\documentclass[letter]{aa}
\usepackage{graphicx}
\usepackage{txfonts}
\usepackage{supertabular}
\renewcommand{\deg}{$^{\circ}$}

\begin{document}

   \title{A stringent upper limit of the PH$_3$ abundance at the cloud top of Venus}

   \author{T. Encrenaz\inst{1}
   \and T. K. Greathouse\inst{2}
   \and E. Marcq\inst{3}
   \and T. Widemann\inst{1}
   \and B. B\'{e}zard\inst{1}
   \and T. Fouchet\inst{1}
    \and R. Giles\inst{2}
    \and H. Sagawa\inst{4}
    \and J. Greaves\inst{5}
    \and C. Sousa-Silva\inst{6}
}

   \institute{LESIA, Observatoire de Paris, PSL Universit\'{e}, CNRS, Sorbonne Universit\'{e}, Universit\'{e} de Paris, 92195 Meudon, France
   \and  SwRI, Div. 15, San Antonio, TX 78228, USA
   \and  LATMOS/IPSL, UVSQ Universit\'{e} Paris-Saclay, Sorbonne Universit\'{e}, CNRS, 78280 Guyancourt, France
   \and Kyoto Sanyo University, Kyoto 603-855, Japan
   \and School of Physics and Astronomy, Cardiff University, Cardiff, UK
   \and Department of Earth, Atmospheric and Planetary Sciences, Massachusetts Institute of Technology, Cambridge, MA, USA
   }

   \offprints{T. Encrenaz}

   \date{Received September 30, 2020, revised October 14, 2020}

   \abstract
{}
{Following the announcement of the detection of phosphine (PH$_3$) in the cloud deck of Venus at millimeter wavelengths, we have searched for other possible signatures of this molecule in the infrared range. }
{Since 2012, we have been observing Venus in the thermal infrared at various wavelengths to monitor the behavior of SO$_2$ and H$_2$O at the cloud top. We have identified a spectral interval recorded in March 2015 around 950 cm$^{-1}$  where a PH$_3$ transition is present.  }
{From the absence of any feature at this frequency, we derive, on the disk-integrated spectrum, a 3-$\sigma$ upper limit of 5 ppbv for the PH$_3$ mixing ratio, assumed to be constant throughout the atmosphere. This limit is 4 times lower than the disk-integrated mixing ratio derived at millimeter wavelengths. }
{Our result brings a strong constraint on the maximum PH$_3$ abundance at the cloud top and in the lower mesosphere of Venus. }


\maketitle

\section{Introduction}
The atmospheric chemistry of Venus is driven by the cycles of water and sulfur dioxide (Krasnopolsky 1986, 2007, 2010; Mills et al. 2007; Zhang et al. 2012). Below the clouds, both species are present with relatively large abundances (about 30 ppmv and 130 ppmv respectively, B\'{e}zard and De Bergh 2007, Marcq et al. 2013) and, at low latitude, are transported upward by Hadley convection. Their abundances drop drastically above the H$_2$SO$_4$ clouds, formed after the SO$_2$  photodissociation and the combination of SO$_3$  with H$_2$O. Their abundances in the mesosphere are  about 1-3 ppmv (Fedorova et al. 2008; Belyaev et al. 2012) and 10-1000 ppbv (Zasova et al. 1993; Marcq et al. 2013, 2020; Vandaele et al. 2017a, b; Encrenaz et al. 2020) respectively.  The water and sulfur dioxide cycles have been extensively monitored over several decades, using Pioneer Venus, the Venera 15 spacecraft, Venus Express and Akatsuki by imaging and spectroscopy in the ultraviolet and infrared ranges. Since 2012, we have been monitoring the abundances of SO$_2$ and H$_2$O (by observing HDO as a proxy) using ground-based imaging spectroscopy in the thermal infrared range with the TEXES (Texas Echelon Cross Echelle Spectrograph) imaging spectrometer, mounted at the Infrared Telescope Facility at Maunakea Observatory (Encrenaz et al. 2020).

In September 2020, the detection of phosphine (PH$_3$) in the cloud decks of Venus was reported on the basis of millimeter heterodyne spectroscopy measurements using both the JCMT and the ALMA facilities (Greaves et al. 2020). This result was a big surprise, as the presence of phosphine is not expected in an oxidized atmosphere, like the ones of the terrestrial planets, if abiotic processes only are considered. 

Since the detection of PH$_3$ was based on the identification of a single transition (at a wavelength of 1.123 mm), we searched for other possible spectral signatures at other wavelengths to confirm and complement the first detection. Phosphine has a very rich infrared spectrum, which ranges from the near-infrared range up to the thermal range. As shown in our previous monitoring of SO$_2$ and H$_2$O, the simultaneous observation of the minor species and weak CO$_2$ lines is needed to infer  the mixing ratio of the species. A good compromise can be found in the 930-960 cm$^{-1}$ region, which includes PH$_3$ transitions with an intensity close to  10$^{-20}$  cm$^{-1}$/molec.cm$^{-2}$ and CO$_2$ lines of various intensities. Within our data set, we found a spectral region around 955 cm$^{-1}$ which includes some relatively strong transitions of PH$_3$ and weak transitions of CO$_2$. We recorded the 951-956 cm$^{-1}$ spectrum with the purpose of analyzing the CO$_2$ hot band for temperature retrieval and for an analysis of the non-LTE effects. Three data sets were obtained in February 2014, March 2015 and January 2016. One PH$_3$ transition is usable in the March 2015 data set and we analyzed it to derive an upper limit of the PH$_3$ mixing ratio; in the two other cases, the PH$_3$ transition fell in the overlap between two consecutive orders of the TEXES instrument (Lacy et al. 2002).

In this paper, we first describe the observations (Section 2). An upper limit of the PH$_3$ abundance at the cloud top is presented in Section 3. Results are discussed in Section 4.

\section{Observations}

TEXES (Texas Echelon Cross Echelle Spectrograph) is an imaging high-resolution thermal infrared spectrograph in operation at the NASA InfraRed telescope Facility at Maunakea Observatory, Hawaii (Lacy et al. 2002), which combines high spectral capabilities (R = 80000 at 7 $\mu$m) and spatial capabilities (around 1 arcsec).

Data were recorded on March 28, 2015, at 01:21:14 UT, between 951 and 956 cm$^{-1}$. The Venus diameter was 14 arcsec and the airmass was 1.016. The Doppler velocity was -11 km/s, corresponding to a Doppler shift of +0.035 cm$^{-1}$ at 950 cm$^{-1}$. The illuminated fraction was 78\% (very similar to the JCMT and ALMA observations reported by Greaves et al. (2020), and the evening terminator was observed. The slit length was 8 arcsec and the slit width was 1.1 arcsec at 950 cm$^{-1}$. We aligned the slit along the North-South celestial axis and we shifted it from west to east, with a step of half the slit width and an integration time of 2 seconds per position, to cover the planet in longitude from limb to limb, and to add a few pixels on the sky beyond each limb for sky subtraction.  As the diameter of Venus was larger than the slit length, we made two scans (North and South) to cover the full latitude range with some overlap around the equator. The total observation time was 18 minutes. The atmospheric transmission is very good around 950 cm$^{-1}$; a single broad feature is observed at 955.25 cm$^{-1}$ (rest frequency) due to terrestrial atmospheric water vapor, which is outside the position of the PH$_3$ transition.

 The TEXES data cubes were calibrated using the standard radiometric method (Lacy et al. 2002, Rohlfs and Wilson 2004). Calibration frames consisting of black chopper blade measurements and sky observations are systematically taken before each observing scan, and the difference (black-sky) is taken as a flat field. If the temperature of the black blade, the telescope and the sky are equal, this method corrects both telescope and atmospheric emissions. 
 
Figure 1 shows two disk-integrated spectra of Venus, with and without the limb contribution, along with a synthetic spectrum corresponding to a PH$_3$  volume mixing ratio of 20 ppbv, constant with altitude (details on the modeling are given in the following section). The strongest PH$_3$ transitions are located between 954 and 956.5 cm$^{-1}$ (Table 1). The doublet at 956.23 cm$^{-1}$ falls within a strong CO$_2$ line, so the useful spectral range is limited to 954-956 cm$^{-1}$. The two strongest PH$_3$ transitions fall in the wings of a strong CO$_2$ line; in addition, the one at 954.445 cm$^{-1}$ coincides with a discontinuity due to an overlap between two consecutive orders. The only usable PH$_3$ transition occurs at 955.23 cm$^{-1}$, and is free of instrumental contamination.
 
 Both TEXES spectra show an emission core at the center of the strong CO$_2$ line at 954.545 cm$^{-1}$.This phenomenon is due to a non-LTE effect in the hot band of CO$_2$ around 955 cm$^{-1}$ (Table 1), which takes place in the upper mesosphere. Individual spectra show that the core emission is especially strong at the limb, but is still slightly present at the center of the disk. In order to minimize at maximum its effect on our analysis (which probes the few kilometers above the cloud top), we have integrated the TEXES data taking into account only the air-masses lower than 1.7, to exclude the limb and the high latitudes contributions. Our summation includes all latitude ranges up to +/- 50 \deg. This spectrum is used for the present analysis. It can be noticed that both TEXES spectra show no significant difference in the vicinity of the PH$_3$ line, with, in both cases, a signal-to-noise close to 1000 (see below).

\begin{figure}
    \includegraphics[width=10cm]{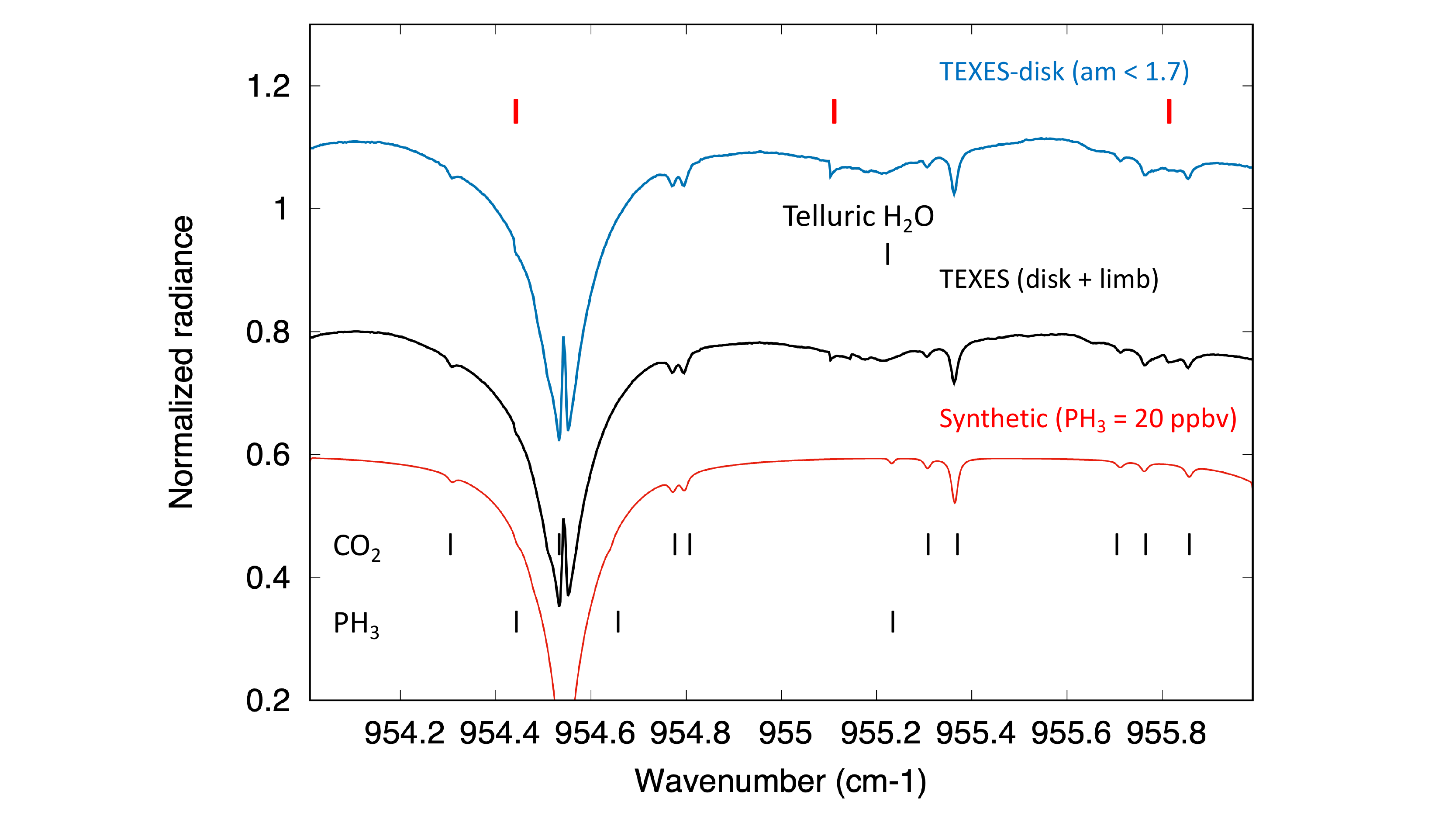}
    \caption{Blue curve: the TEXES spectrum recorded on March 28, 2015 between 954 and 956 cm$^{-1}$, integrated for all air-masses lower than 1.7. Black curve (shifted by -0.2 for clarity): the TEXES disk-integrated spectrum of Venus, extracted from the same data set,  including the limb contribution. Red curve (shifted by -0.4 for clarity): synthetic spectrum including CO$_2$ and PH$_3$ with a volume mixing ratio of 20 ppbv, constant with altitude, calculated for an airmass of 1.15 (30 \deg latitude). The broad absorption feature around 955.215 cm$^{-1}$ is due to a telluric water vapor line. The red ticks at the top of the figure indicate the discontinuities due to an overlap between two consecutive orders.}
\end{figure}

\begin{table*}
\begin{tabular}{|l|l|l|l|l|l|l|}
\hline Molecule.  & Wavenumber   &   Band identification & Line identification & Intensity &  Energy   &   Broad. coef.   \\
                           & cm$^{-1}$              &                       &                             & cm molec$^{-1}$    &   cm$^{-1}$       &    cm$^{-1}$  atm$^{-1}$          \\ \hline
PH$_3$               & 954.44508            &      0100 0000 & 3 0 0A+  4 0 0A+ &1.343 $\times$ 10$^{-20}$ &  88.9959        &      0.1014               \\
PH$_3$               & 954.64090            &      0100 0000 & 3 1 0E  4 1 0E    &1.262 $\times$ 10$^{-20}$ &  88.4658        &      0.1026                   \\
PH$_3$               & 955.23186           &      0100 0000 &  3 2 0E  4 2 0E      &1.02 $\times$ 10$^{-20}$ &  86.8739         &      0.1076                    \\
PH$_3$               & 956.2288             &      0100 0000 & 3 3 0A-  4 3 0A-    &6.09 $\times$ 10$^{-21}$ &  84.2149       &      0.1078                \\
PH$_3$               & 956.2288             &      0100 0000 & 3 3 0A-  4 3 0A+    &6.09 $\times$ 10$^{-21}$ &  84.2149       &      0.1078                    \\
CO$_2$               & 955.306951         &    000 11 100 01 & P11E                 & 6.663 $\times$ 10$^{-27}$ & 1426.0242 &      0.1076                   \\
 \hline 

\end{tabular}
\caption{Spectroscopic parameters of the strongest PH$_3$ transitions between 949 and 959 cm$^{-1}$, extracted from the GEISA-2015 database. The CO$_2$ transition used in the calculations is also added.  The line intensities correspond to a temperature of 296 K. The broadening coefficients (HWHM) refer to the broadening  by CO$_2$. }
\end{table*}  
  
\section{PH$_3$ upper limit}

Synthetic spectra of PH$_3$ in the atmosphere of Venus were calculated using the radiative transfer code that we apply for monitoring SO$_2$ and HDO at the cloud top (Encrenaz et al. 2016, 2019). This line-by-line code calculates the outgoing flux using an integration over 175 atmospheric levels separated by 1 km. The cloud top is defined by a blackbody at a temperature of 235 K and a pressure of 150 mbars. The thermal profile is the same as used in Encrenaz et al. (2016). The temperature is 210 K at 11 km  above the cloud top (P = 10 mbars), 187 K at 21 km above this level (P = 1 mbar) and 180 K for altitudes higher than 28 km above this level ( P lower than 0.2 mbar). The penetration level in the mid-infrared is governed by the extinction cross-section of the H$_2$SO$_4$ particles which constitute the upper cloud deck. As illustrated by Zasova et al. (1993), the values of this coefficient are very similar at 950 cm$^{-1}$ and at 1350 cm$^{-1}$, where our analysis of SO$_2$ and HDO was performed.  The atmospheric model used in our previous analyses is thus suited for the present study.

The spectroscopic data for PH$_3$ were extracted from the GEISA-2015 database (Jacquinet-Husson et al. 2016). For the broadening coefficients PH$_3$-CO$_2$, in the lack of more precise information, we assumed, as we did for SO$_2$ and HDO (Encrenaz et al. 2016), an increase by a factor of 1.4 with respect to the air-broadening coefficients (Nakazawa and Tanaka 1982). As a verification, we estimated independently the broadening coefficient of our PH$_3$ line at 955.231 cm$^{-1}$ by using NH$_3$ as an analog, as done by Greaves et al. (2020) for the PH$_3$ millimeter line. The air-broadened HWHM (half-width at half-maximum) of our PH$_3$ transition at 954.232 cm$^{-1}$ is 0.0744 cm$^{-1}$ atm$^{-1}$. The NH$_3$ transition with the same quantum numbers as our PH$_3$ transition ($\nu$$_2$, J = 4, K = 3) has an air-broadened HWHM of 0.1017 cm$^{-1}$. The HWHM(PH$_3$)/HWHM(NH$_3$) is 0.0744/0.1017 = 0.732. Using the polynomial described by Wilzewski  et al. (2016), we infer a CO$_2$-broadened HWHM of 0.1537 cm$^{-1}$ atm$^{-1}$ for the NH$_3$ transition. We thus derive, for the CO$_2$ HWHM of our PH$_3$ transition, a value of 0.157x0.732 = 0.112 cm$^{-1}$ atm$^{-1}$, very close to the value listed in Table 1.
 
Figure 2 shows an enlargement of the disk-integrated TEXES spectrum of March 28, 2015 (without the limb contribution), in the vicinity of the PH$_3$ transition at 955.23 cm$^{-1}$, compared with synthetic spectra of CO$_2$ and PH$_3$, for volume mixing ratios of 5, 10 and 20 ppbv, constant with altitude. As mentioned above, the slope of the observed spectrum,  around the position of the PH$_3$ line, is due to the presence of a broad H$_2$O telluric line. It can be seen that there is no trace of PH$_3$ absorption in the TEXES spectrum.

In order to derive an upper limit for  the PH$_3$ abundance at the cloud top of Venus, we have assumed, as in the case of SO$_2$ and HDO, that the PH$_3$/CO$_2$ line depth ratio varies linearly with the PH$_3$ volume mixing ratio. In the case of Mars, we have shown that this method is valid for line depths weaker than ten percent for deriving H$_2$O$_2$ and HDO volume mixing ratios from H$_2$O$_2$/CO$_2$ and HDO/CO$_2$ line depth ratios; the uncertainty is a few percent for  air-masses lower than 2 (Encrenaz et al. 2008, 2015a). In the case of Venus, we have shown that, in the 1350 cm$^{-1}$ range, for SO$_2$ and HDO lines weaker than ten percent in depth, the linearity is verified with an uncertainty of 7 percent for an air-mass of 1.4  (Encrenaz et al. 2012). 

We estimated the peak-to-peak (3-$\sigma$) variations of the TEXES continuum in the vicinity of the PH$_3$ transition, between 955.20 and 955.27 cm$^{-1}$. We found a value of 0.001, corresponding to a S/N of about 1000. We checked that this high signal-to-noise ratio is actually achieved between the lines over the whole range of the spectrum (Figure 1). The CO$_2$ line depth in the TEXES spectrum is 0.01. The PH$_3$/CO$_2$ line depth  ratio in the TEXES spectrum is thus lower than 0.10, while this ratio, in the synthetic spectrum, is 0.11 for a PH$_3$ volume mixing ratio of 5 ppbv. We  thus infer a 3-$\sigma$ upper limit of 5 ppbv for the PH$_3$ mixing ratio at the cloud top of Venus.

\begin{figure}
    \includegraphics[width=10cm]{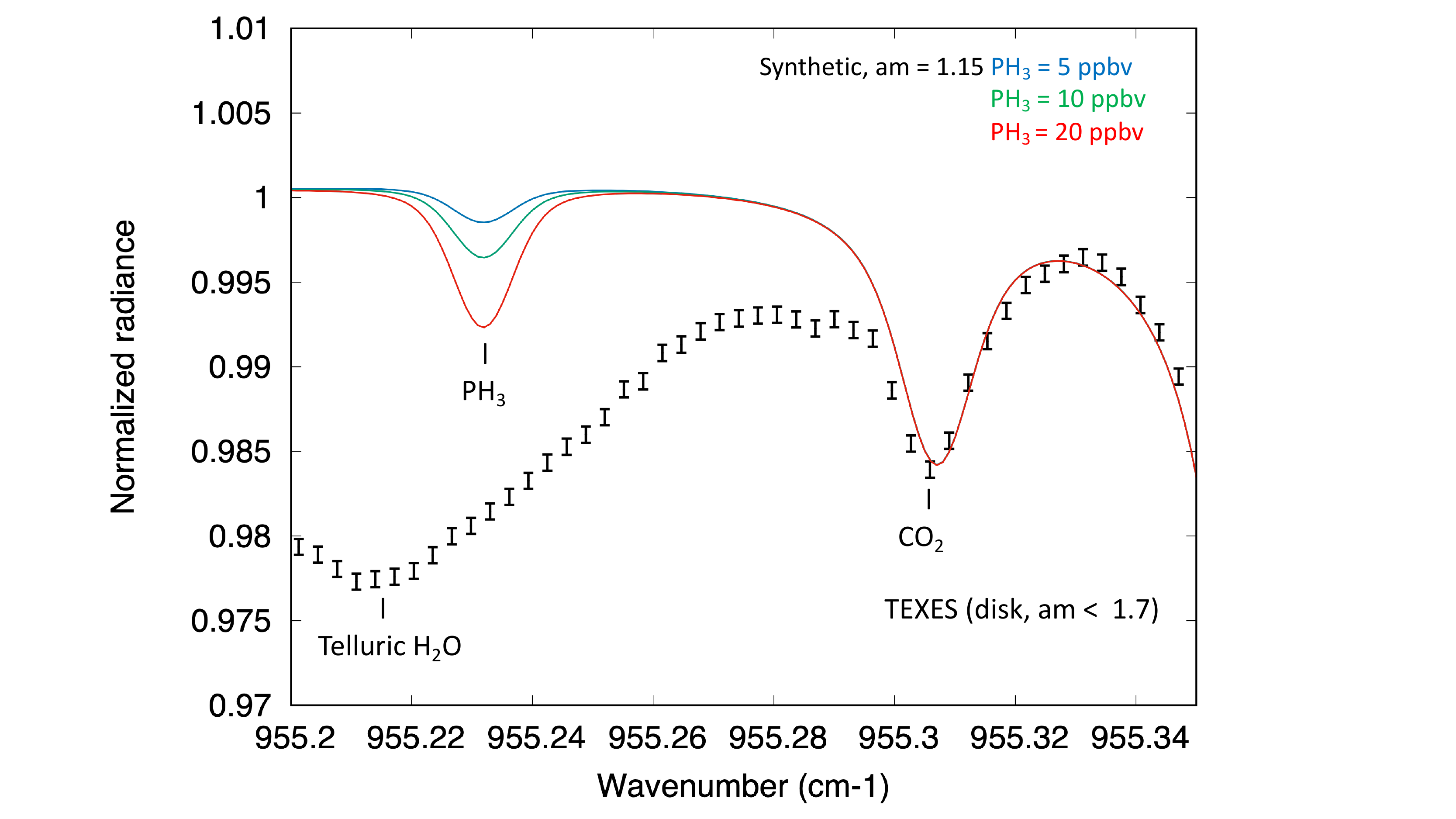}
    \caption{Black error bars (3-$\sigma$): The disk-integrated spectrum of Venus (without the limb contribution) between 955.20 and 955.35 cm$^{-1}$ recorded on March 28, 2015.  Models: CO$_2$ and  PH$_3$ (blue: 5 ppbv; green: 10 ppbv; red: 20 ppbv), calculated for an airmass of 1.15 and for a constant PH$_3$ volume mixing ratio throughout the mesosphere. The slope of the TEXES spectrum is due to the broad telluric H$_2$O line centered at 955.215 cm$^{-1}$.}
    \end{figure}
    
 As a next step, we searched for possible local variations of the signal at the position of the PH$_3$ line over the disk of Venus. To do so, we used the same method as in the case of our SO$_2$ and HDO maps, which are derived from the SO$_2$/CO$_2$ and HDO/CO$_2$ line depth ratios, respectively.  We defined the depth of the pseudo-PH$_3$ line by taking the signal at the line center, divided by the mean value of the continuum on each side of this position. We measured, in the same way, the depth of the CO$_2$ line at 955.3069 cm$^{-1}$, and we took the ratio of both quantities. The result is shown in Figure 3.  It can be seen that the PH$_3$/CO$_2$ line depth ratio is always between +0.05 and -0.05, except at high southern latitudes where the ratio is close to 0.10. These high values are meaningless, because, at these latitudes, the CO$_2$ line depth is zero or negative, indicating a different behavior of the temperature profile at the level of the southern polar collar.  The upper limit for the PH$_3$ mixing ratio inferred from Figure 4 is 3.5 ppbv. The fact that this value is lower than our disk-integrated limit is due to the fact that, in the mapping process, the fluxes are averaged over 3 pixels, both at the line center and on each side of the line. This result confirms our 3-$\sigma$ upper limit of 5 ppbv.

\begin{figure}
    \includegraphics[width=12cm,angle=270]{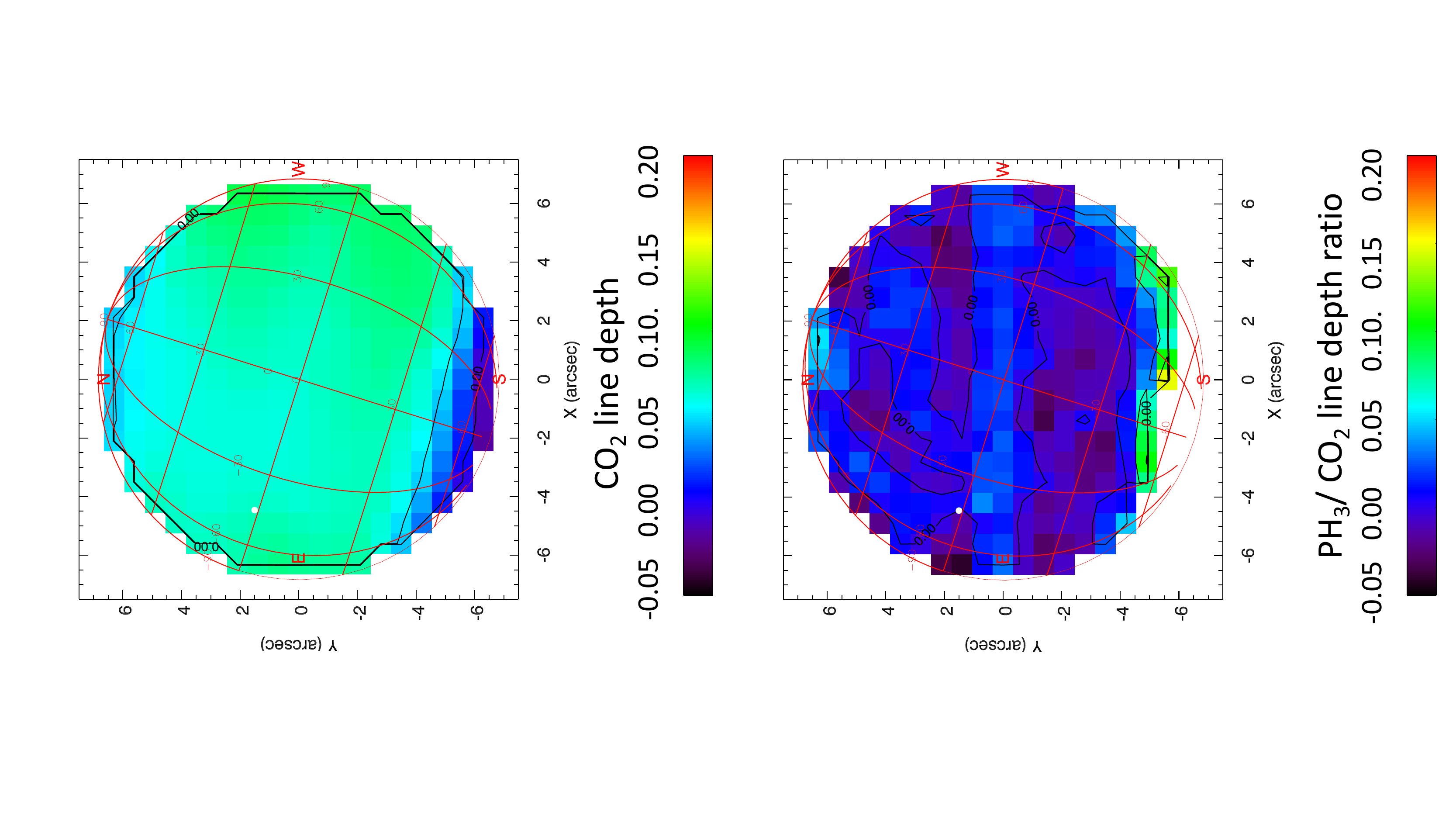}
    \caption{Top: map of the line depth of the weak CO$_2$ transition at 955.3069 cm$^{-1}$, corresponding to the observations of March 28, 2015  shown in Figures 2 and 3. Bottom: map of the PH$_3$/CO$_2$ line depth ratio. The sub-solar point is shown as a white dot. The negative values in the map of the CO$_2$ line depth (top) indicate a different behavior of the temperature profile at the level of the polar collar, at high southern latitudes. }
\end{figure}

\section{Discussion}

Our upper limit of PH$_3$ is to be compared with the detection of phosphine in the millimeter range. Our upper limit is not compatible with a constant mixing ratio of 20 ppb throughout the mesosphere, as announced by Greaves et al. (2020). We might wonder if the difference between the two results could be due to the altitude difference of the regions probed at both wavelengths. Indeed, the continuum thermal emission at millimeter wavelengths probes an altitude around 55 km, i.e. within the clouds. This level is close to the one probed at 19 $\mu$m, in the $\nu$$_2$ band of SO$_2$ (in our model, z = 57 km, P = 250 mb, T = 241 K). The TEXES measurements at 10-$\mu$m, like the 7-$\mu$m observations of the $\nu$$_3$ SO$_2$ band, probe the cloud top (in our present model, located at a pressure of 150 mbar and a temperature of 235 K, corresponding to an altitude of about 60 km; Encrenaz et al. 2016, 2019). 

So we might wonder if  a PH$_3$ vertical profile enriched in the upper cloud at 55 km and depleted above 60 km would resolve the discrepancy. However it is not the case, because the core of millimeter line is very narrow (less than about 20 MHz). If the PH$_3$ millimeter line was formed within the clouds, at a pressure level of 100 mbars or more, its HWHM would be at least 0.01 cm$^{-1}$, i.e. 300 MHz. Such a broad line would not be observable by heterodyne spectroscopy. This implies that the millimeter line observed by Greaves et. al. (2020) must be formed relatively high in the mesosphere. An analogy can be drawn with the millimeter lines of SO$_2$ and SO, formed at about 80 km (Sandor et al., 2010), whose presence requires the existence in the upper mesosphere of a sulfur reservoir, still not identified. On the other hand, it must be reminded that the TEXES data, when using very weak lines, probe only the few kilometers above the cloud top; they are not sensitive to levels higher than about ten kilometers above this level (Encrenaz et al. 2013). The analysis of the constraints provided by the ALMA and TEXES observations on the vertical profile of the observed millimeter line will be the subject of a forthcoming publication. We also note that the narrow width of the PH$_3$ 1-0 line could possibly be attributed to the removal of the line wings during the process of removing the baseline ripples, which could narrow the line core  (Greaves et al 2020, in preparation).

Another explanation for the discrepancy could be the variability of phosphine. Mesospheric sulfur species observed in the millimeter range (SO, SO$_2$) are known to vary as a function of time and space (Sandor et al. 2010; Encrenaz et al. 2015b; Piccialli et al. 2017). In any case, the detection of at least one other PH$_3$ transition, in the infrared or in the millimeter/sub-millimeter range, is definitely needed to confirm the PH$_3$ detection in Venus.

\begin{acknowledgements}
TE, TKG et RG were visiting astronomers at the NASA Infrared Telescope Facility, which is operated by the University of Hawaii under Cooperative Agreement no. NNX-08AE38A with the National Aeronautics and Space Administration, Science Mission Directorate, Planetary Astronomy Program. We wish to thank the IRTF staff for the support of TEXES observations. This work was supported by the Programme National de Plan\'{e}tologie (PNP) of CNRS/INSU, co-funded by CNES. TKG acknowledges support of NASA Grant NNX14AG34G. TE and BB acknowledge support from CNRS. TF acknowledges support from Sorbonne Universit\'{e}. TW acknowledges support from the University of Versailles-Saint-Quentin and the European Commission Framework Program FP7 under Grant Agreement 606798 (Project EuroVenus).  We thank E. Lellouch and J. Lequeux for helpful comments regarding this letter.

\end{acknowledgements}

\end{document}